\documentclass[12pt]{article}

\usepackage{moreverb,url}

\usepackage{authblk}

\usepackage[colorlinks,bookmarksopen,bookmarksnumbered,citecolor=red,urlcolor=red]{hyperref}

\newcommand\BibTeX{{\rmfamily B\kern-.05em \textsc{i\kern-.025em b}\kern-.08em
T\kern-.1667em\lower.7ex\hbox{E}\kern-.125emX}}

\pdfoutput=1

\usepackage[super,comma]{natbib}
\usepackage{amsmath}
\usepackage{amsthm}
\usepackage{amsfonts,bm}
\usepackage{a4}
\usepackage{graphicx}
\usepackage[utf8]{inputenc}
\usepackage[english]{babel}

\usepackage{authblk}

\usepackage{enumerate}
\usepackage{enumitem}
\usepackage{float}
\usepackage{newcent}
\usepackage{mathrsfs}

\usepackage{float}

\usepackage[bottom=2.5cm]{geometry}
\usepackage{cancel}
\usepackage{listings,color,verbatim}
\usepackage{fancyvrb}
\usepackage{xcolor}
\usepackage{pdfpages}

\usepackage{framed,color}

\usepackage{multirow}
\usepackage{amstext}
\usepackage{array}
\newcolumntype{C}{>{$}c<{$}}

\begin{document}

\title{A Bayesian multivariate spatial approach for illness-death survival models.}

\author[1*]{\large Fran Llopis-Cardona}
\author[2]{Carmen Armero}
\author[1,3,4]{Gabriel Sanf\'elix-Gimeno}
\affil[1]{\small Health Services Research Unit, Foundation for the Promotion of Health
and Biomedical Research of Valencia Region (FISABIO), Valencia, Spain.}
\affil[2]{\small Department of Statistics and Operations Research. Universitat de Val\`encia, Spain.}
\affil[3]{\small Red de Investigaci\'on en Servicios de Salud en Enfermedades Cr\'onicas (REDISSEC), Valencia, Spain.}
\affil[4]{\small Network for Research on Chronicity, Primary Care, and Health Promotion (RICAPPS).}
\affil[*]{Corresponding author. F. Llopis-Cardona. E-mail: \texttt{llopis\_fracar@gva.es}}

\date{} 

\maketitle

\begin{abstract}
Illness-death models are a class of stochastic models inside the multi-state framework. In those models, individuals are allowed to move over time between different states related to illness and death. They are of special interest when working with non-terminal diseases, as they not only consider the competing risk of death but also allow to study progression from illness to death. The intensity of each transition can be modelled including both fixed and random effects of covariates. In particular, spatially structured random effects or their multivariate versions can be used to assess spatial differences between regions and among transitions. We propose a Bayesian methodological framework based on an illness-death model with a multivariate Leroux prior for the random effects. We apply this model to a cohort study regarding progression after osteoporotic hip fracture in elderly patients. From this spatial illness-death model we assess the geographical variation in risks, cumulative incidences, and transition probabilities related to recurrent hip fracture and death. Bayesian inference is done via the integrated nested Laplace approximation (INLA).
\end{abstract}


\maketitle

\section{Introduction}


Multi-state models are stochastic models which generalize a wide class of survival scenarios, from unidimensional survival models to multi-event models such as competing risks models or repeated events \cite{andersen competing, kneib}. In the multi-state framework, events  are the states of the process, and  their respective occurrences are transitions between  the state of departure and the state of interest.   The uncertainty associated to transitions is modelled via transition probabilities or, equivalently, transition intensities.  The latter are analogue to hazard functions in the field of survival analysis. Multi-state models are especially useful in medical research because they provide  a natural setting for dealing with   the natural history of complex diseases \cite{Le}.

The so-called illness-death model \cite{andersen} is one of the simplest and most studied multi-state models. It has three states:  an initial state, an illness-related state and a death state. The process starts in the initial state from which it can progress to the illness transient state or to death, which is an absorbent state. Death is also accessible from the illness state.  This model is particularly useful to the study of chronic diseases \cite{vejakama}, cancer progression \cite{armero illness-death} or cardiovascular diseases \cite{kuhn} in which there is a considerable risk of death over time.

The Cox proportional hazards model \cite{cox prop} is the most popular regression tool in the survival framework to model hazard functions associated to survival times. It expresses hazard functions as the product of a time-dependent baseline hazard function and the exponential of a regression term including covariates and latent elements. The popularity of this model is primarily due two reasons. First, due to  the interpretability of hazard ratios to evaluate differences in the risk of the event of interest among the different covariate levels.  Second, because under the frequentist paradigm that objective does not require to make any assumptions about the baseline hazard function. From a Bayesian reasoning, a model for the baseline risk function needs to be specified \cite{christensen}, either parametrically or semi-parametrically  \cite{lazaro}. In particular, when we work only with covariates and use a Weibull baseline risk function, the overall risk function  will also be Weibull.   This property is the basis of the corespondence between  the Weibull Cox proportional hazards model and the Weibull accelerated failure time (AFT) regression model \cite{collett}.

The Bayesian paradigm provides a flexible framework for statistical inferences and generation of knowledge. Under this framework any measure of interest is subject to uncertainty: not only random variables  but also parameters, hypothesis, models, etc. On the other hand, the Bayesian inferential process via the Bayes’ theorem allows to sequentially update previous knowledge of those measures using new information. The procedure is conceptually simple: the elicitation of a  prior distribution  for all uncertainties in the model, the computation of the likelihood function for the data obtained, and the estimation of the posterior distribution which   updates the relevant knowledge. This posterior is the starting point to approximate the posterior distribution of any measure of interest, such as sojourn times, transition and occupation probabilities, and cumulative incidence functions.

Regression survival models can include not only covariates  but also latent effects  that account for some  non-explained heterogeneity between groups of the target population.    In particular, the existence of differences among spatial regions is especially common in epidemiological studies. Uncontrolled risk factors may be relevant to explain high or low risks of disease in some regions, leading to this heterogeneity. We focus in this paper on lattice data, i.e. data for a finite number of sub-regions of a larger one. Often neighbouring regions can be expected to be similar and thus random effects with a spatial correlation structure can be assumed.

There is  a plenty of models for assessing  spatial correlation in the statistical literature. Conditional autoregressive (CAR) models \cite{besag} and their variants based on a neighbourhood definition of the correlation have been widely used in disease mapping. In particular, the model proposed by Besag, York and Mollié (BYM) in 1991 \cite{bym} has been postulated as the main choice over the past decades to deal with counts assuming a Poisson process. It is defined by means of two random effects, the first based on the neighbourhood structure and thus summarizing the spatial correlation between regions, and a second unstructured accounting for heterogeneity among regions. Leroux et al. (1999) \cite{leroux} propose an alternative specification for the precision matrix of the spatially distributed random effects that better distinguishes between spatial dependence and dispersion effects. Under this model random effects are defined as a mixture of independent and spatially correlated scenarios. Some authors assessed the behaviour of spatial models inside the survival framework such as Banerjee, Wall and Carlin (2003) \cite{banerjee frailty}, comparing different models without random effects (usually referred as frailties in the survival setting), with non-spatial frailties and with a CAR frailty.

Regarding illness-death models, not only spatial correlation can be modelled, but also correlation between the three transitions, resulting in a multivariate model for random effects. In this regard, Carlin and Banerjee (2003) \cite{carlin multivariate} proposed a multivariate CAR model for spatially correlated survival times. However, and despite its interest there are few studies considering spatial components in the illness-death model framework. The most remarkable research work in this direction is Nathoo and Dean (2007) \cite{nathoo}  in which various structures for  region-specific random effects are proposed, with especial attention  to the comparison of different baseline functions such as Weibull distributions, piecewise-exponential forms and cubic B-splines.

We propose a Bayesian methodological framework to deal with spatially-correlated random effects within the illness-death scenario. In particular, a multivariate version of the Leroux model is used to jointly model that spatial correlation  as well as  the correlation between  the transition survival times. The Bayesian procedure involving the approximation of   the relevant  posterior distribution is done via the integrated nested Laplace approximation (INLA), which in general provides accurate estimations and reduces the computational time compared to Markov chain Monte Carlo (MCMC) methods. In the context of the proposed methodological framework, the computation of posterior outcomes such as  sojourn times, transition and occupation probabilities and cumulative incidence functions   results in   natural outcomes. Moreover, those quantities can be mapped providing rich information about the spatial distribution of illness and death in terms of probabilities that may have important clinical implications as interpreted by clinicians and epidemiologists. We apply this model to a real-world study involving recurrent hip fracture in old people. Data come from the PREV2FO cohort of patients from the Comunitat Valenciana (Spain) aged 65 and over who have been discharged from hospital after a hip fracture. They include  individual baseline information of those patients  and their progression over time. In addition to these individual characteristics, Health Areas where patients belong to are included to assess geographical differences in their corresponding risks and transition probabilities.

This paper is structured as follows: illness-death model is presented in Section \ref{sec:illness-death model}; the methodological framework regarding  the proposed spatial illness-death model is described in Section \ref{sec:bayes-illness-death model}, including the sampling model in Subsection \ref{subsec:sampling}, Bayesian inference and prior specification in Subsection \ref{subsec:prior}, and some posterior outcomes in Subsection \ref{subsec:outcomes}. Section \ref{sec:appli} includes the analysis of the study of recurrent hip fracture, in particular, Subsection \ref{subsec:prev2fo} presents the PREV2FO cohort, Subsection \ref{subsec:posterior} includes some relevant results regarding posterior inference of the parameters of the model, and Subsection \ref{subsec:measures} includes the results regarding relevant measures of the process such as cumulative incidences and transition probabilities. Finally, we present a discussion in Section \ref{sec:discu}.

\section{Illness-death models}\label{sec:illness-death model}

\begin{figure}[H]
  \centering
  \includegraphics[width=3cm]{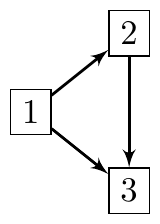}
  \caption{Illness-death model with initial state (1), transient illness state (2) and death state (3). Arrows between the states represent the corresponding transitions.  }\label{diag}
\end{figure}

Illness-death models are the  most popular multi-state models \cite{andersen}. In their simplest version they comprise three states: an initial state (1), an illness-related state (2), and a death state (3). The death state is absorbent and accessible directly from the initial state or through the intermediate and transient state  defined by the illness. Figure \ref{diag} depicts the model, including transitions between the states.

From a probabilistic framework, an illness-death model is defined as a stochastic process $\{Z(t), t>0\}$ in continuous time $t$  which takes as values the possible states where individuals can be. In particular,  $S=\{1,2,3\}$ is the state space  and $C=\{1\rightarrow2, \,1\rightarrow3, \,2\rightarrow3\}$  the set which includes all possible transitions of the process. We assume a semi-Markovian \cite{meira smmr} structure of the process whose  evolution  from the initial state to  disease or to death only depends on the history of the process through the current state; but the transition from   disease to   death will depend not only on the present situation of the process but also on how long it has been in the initial state before jumping to the disease state.

The random behaviour of our illness-death model is determined by the initial distribution of the process, $\textrm{P}(Z(0)=i)$ $\forall i \in S$ (usually $\textrm{P}(Z(0)=1)=1$ because it is assumed that at $t=0$ individuals are in state $1$),  and the so-called transition probabilities between states defined as:
\begin{align}
&p_{1j}(s,t)= \textrm{P}(Z(t)=j\mid Z(s)=1), \quad s \leq t, \,\,j=2,3, \label{trans_prob1}\\
&p_{23}(s,t \mid t_{12})=\textrm{P}(Z(t)=j\mid Z(s)=1, T_{12}=t_{12}), \quad s \leq t_{12} \leq t, \nonumber
\end{align}

\noindent where $T_{12}$ is the time  the process spends in state $1$ before entering into state $2$.
 Transition probabilities provide intuitive and easily interpretable information about the problem of interest but are difficult to model.  For this reason, the statistical modelling usually relies on transition intensities, which are much less intuitive but easier to model. They   account for the instantaneous hazard of progression to state $j$ conditional on the current state $i$ as follows
\begin{align}
h_{ij}(t) = & \lim_{\Delta t \rightarrow 0}\dfrac{p_{ij}(t, t +\Delta t)}{\Delta t}. \label{trans_intens1}
\end{align}
Notice that transition $2\rightarrow3$ needs to add the condition $T_{12}=t_{12}$ to the subsequent transition intensity.

In the case of our illness-death model, transition probabilities are computed from transition intensities as indicated below  (Armero \textit{et al.}, 2016) \cite{armero illness-death}:
\begin{align}
&p_{11}(s,t)=\exp{\Big\{-\int_s^t \,(h_{12}(u)+h_{13}(u)) \mbox{d}u \Big\}}\nonumber\\ \label{trans_prob2}
&p_{22}(s,t \mid t_{12})=\exp{\Big\{-\int_s^t h_{23}(u-t_{12}|t_{12})\mbox{d}u \Big\}} \nonumber\\
&p_{12}(s,t )=\int_s^t p_{11}(s,u) h_{12}(u)p_{22}(u,t| u) \mbox{d}u \nonumber\\
&p_{13}(s,t)=1-p_{11}(s,t)-p_{12}(s,t)\nonumber\\
&p_{23}(s,t \mid t_{12})=1-p_{22}(s,t)\nonumber\\
&p_{33}(s,t)=1,
\end{align}

The equivalence between some concepts from the world of stochastic processes and of survival analysis is quite natural for the given setting since the transition time from $i$ to $j$ can be seen as the survival time between the initiating event $i$ and the  entrance in the state of interest $j$, $T_{ij}$.  In this regard,
 transition intensities     in the stochastic framework (\ref{trans_intens1})  are equivalent to hazard functions in  the survival setting as follows
 \begin{align}
h_{ij}(t) = & \lim_{\Delta t \rightarrow 0}\dfrac{\textrm{P}(t\leq T_{ij}< t+\Delta t \mid T_{ij} \geq t)}{\Delta t}. \label{trans_intens2}
 \end{align}

Transition intensities can be naturally modelled using Cox proportional hazard models \cite{cox prop}, which expresses hazard functions by means of the product of a baseline hazard function and an exponential  regression term with covariates, random effects or   any other element that can provide information on the variable of interest.

\section{Bayesian spatial illness-death modelling}\label{sec:bayes-illness-death model}

We propose a Bayesian spatial   illness-death model for a   finite spatial lattice data   with a set of local neighbourhoods defined by geographic vicinity between the sites of the target region. This model includes the joint modelling of the three relevant survival times of the illness-death model associated to each site as well as a spatial structure, based on the Leroux model \cite{leroux}, which connects the survival process of the different sites of the spatial domain. The notation of the Bayesian content that we will introduce from now on considers all probabilities and derived concepts to be conditional because all the parameters and hyperparameters on which they depend have probability distributions.

\subsection{Sampling model}\label{subsec:sampling}

Let $h_{ij}^{(k)}(t\mid \boldsymbol{\theta})$ denote the conditional hazard function  for survival time $T_{ij}$ associated to transition $i\rightarrow j \in C$ at time $t$ for an individual from region $k$, $k=1, \ldots, K$  which we express through the Cox model \cite{cox prop}
\begin{align}
&h_{ij}^{(k)}(t \mid \boldsymbol{\theta}, \boldsymbol{\psi}) = h_{ij,0}(t\mid \boldsymbol{\theta})\,\exp\{\eta_{ij}^{(k)}\}, \label{eqn:Cox}\\
&\eta_{ij}^{(k)}=\boldsymbol{x}^{\prime} \boldsymbol{\beta}_{ij}  + b_{ij}^{(k)}, \nonumber
\end{align}
\noindent where $\boldsymbol{\theta}$ is the vector of parameters  and hyperparameters of the model, $\boldsymbol{\psi}$ the vector of all random effects, $h_{ij,0}(t\mid \boldsymbol{\theta})$  a baseline hazard function,   and $\eta_{ij}$  a regression term defined in term of covariates $\boldsymbol{x}=(x_1, \ldots, x_L)^{\prime}$, a vector of regression coefficients $\boldsymbol{\beta}_{ij}=(\beta_{ij,1}, \ldots, \beta_{ij,L})^{\prime}$ and a random effect  $b_{ij}^{(k)}$ associated with transition   $i\rightarrow j$  in region $k$. Baseline hazard functions can be approached in several ways. From parametric models such as the Weibull, undoubtedly the traditional and most widely used model in biometric applications, to more flexible modelling such as piecewise constant functions or B-splines (Ibrahim, Chen and Sinha, 2001) \cite{ibrahim}. In our case, we propose a full parametric approach via Weibull baseline hazard functions defined as $h_{ij,0}(t \mid \boldsymbol{\theta})=\alpha_{ij}\lambda_{ij}t^{\alpha_{ij}-1}$.

Random effects $b_{ij}^{(k)}$ in (\ref{eqn:Cox}) depend on the different model transitions as well as of the different sites in the target region. Let $\boldsymbol{B}$ be a matrix which comprises all random effects:

\begin{equation}
\boldsymbol{\mathrm{B}}=\left(
  \begin{array}{ccc}
    b_{12}^{(1)} & \quad b_{13}^{(1)} & \quad b_{23}^{(1)} \\
    \vdots       & \quad \vdots       & \quad \vdots \\
    b_{12}^{(k)} & \quad b_{13}^{(k)} & \quad b_{23}^{(k)} \\
    \vdots       & \quad \vdots       & \quad \vdots \\
    b_{12}^{(K)} & \quad b_{13}^{(K)} & \quad b_{23}^{(K)} \\
  \end{array}
\right),
\end{equation}

 \noindent $\boldsymbol{\mathrm{B}}(,i)$   the $i$-th column of the matrix $\boldsymbol{\mathrm{B}}$, and   $vec(\boldsymbol{\mathrm{B}})=[\boldsymbol{\mathrm{B}}(,1)^{\prime}, \boldsymbol{\mathrm{B}}(,2)^{\prime}, \boldsymbol{\mathrm{B}}(,3)^{\prime}]^{\prime}$   a column vector $(3K \times 1)$ including each of the columns of matrix $\boldsymbol{\mathrm{B}}$. We assume a conditional multivariate Gaussian Markov random field for the random effects   $vec(\boldsymbol{\mathrm{B}})$ with a mean vector whose elements are all zero and a matrix of variances-covariances $\Sigma$
\begin{equation}
(vec(\boldsymbol{\mathrm{B}})\mid \boldsymbol \Sigma) \sim N_{3K}(0, \boldsymbol{\Sigma}).
\end{equation}
\noindent It is worth mentioning that $vec(\boldsymbol{\mathrm{B}})$ is precisely the set of random effects that we have generically represented as $\boldsymbol \psi$ above.

The structure of $\boldsymbol \Sigma$ includes both  multivariate dependence   between the illness-death transitions of a given site (between columns of $\boldsymbol{B}$)  and   spatial dependence for each transition of the model (within columns of $\boldsymbol{B}$) in the form
\begin{equation}
\boldsymbol{\Sigma}= \boldsymbol{\Sigma}_{between} \otimes \boldsymbol{\Sigma}_{within},
\end{equation}

\noindent where $\otimes$ represents the Kronecker product.

Models for the spatial   variability  $\boldsymbol{\Sigma}_{within}$   have a wider range of options: from the simplest independent scenario to the conditional autoregressive (CAR) models \cite{besag} and their variants (intrinsic CAR, proper CAR, Besag York \& Mollié, Leroux model). The proposal by Besag, York \& Mollié (BYM) \cite{bym} has been widely used for disease mapping in the epidemiological literature over the past decades. Part of its popularity remains in its interpretability, as it   consists of   a random effect, which considers spatial correlation between regions according to a neighbourhood structure, and an unstructured random effect accounting for heterogeneity among regions. With the BYM model, however, only the sum of both sets of random effects is identifiable, failing thus to identify both random effects separately \cite{eberly}. The Leroux model \cite{leroux} circumvents this problem since there is only one set of random effects defined in terms of a mixture of independent and spatially-dependent elements that allows to assess the intensity of them as follows
\begin{equation}
\boldsymbol{\Sigma}_{within}=\Big(\tau\,[(1-\gamma)\boldsymbol{\mathrm{I}}+\gamma(\boldsymbol{\mathrm{D}}-\boldsymbol{\mathrm{W}})]\Big)^{-1},
\end{equation}

\noindent where $\tau$ is a dispersion hyperparameter, $\boldsymbol{\mathrm{I}}$ the identity matrix, $\boldsymbol{\mathrm{D}}$ a diagonal matrix    whose non-zero elements   on the diagonal are the number of neighbours in the corresponding  site, $\boldsymbol{\mathrm{W}}$ an adjacency matrix, i.e, $\boldsymbol{\mathrm{W}}_{kl}=1$ if sites $k$ and $l$ are neighbours, $k\neq l$, and 0 otherwise, and hyperparameter $\gamma \in [0, 1]$ determines how matrices $\boldsymbol{\mathrm{I}}$ and $\boldsymbol{\mathrm{D}}-\boldsymbol{\mathrm{W}}$
are combined. A value of $\gamma=0$ simplifies  to an independent random effects model without spatial patterns, whilst $\gamma=1$ corresponds to an intrinsic CAR model.

We model the variance-covariance matrix  $\boldsymbol{\Sigma}_{between}$ of the times between the three  transitions as
\begin{equation}
\boldsymbol{\Sigma}_{between}=\left(
  \begin{array}{ccc}
    \dfrac{1}{\tau_{12}}                                     & \quad \dfrac{\rho_{(12)(13)}}{\sqrt{\tau_{12} \tau_{13}}}        & \quad \dfrac{\rho_{(12)(23)}}{\sqrt{\tau_{12} \tau_{23}}}  \\ \\
    \dfrac{\rho_{(12)(13)}}{\sqrt{\tau_{12} \tau_{13}}}      & \quad \dfrac{1}{\tau_{13}}                                       & \quad \dfrac{\rho_{(13)(23)}}{\sqrt{\tau_{13} \tau_{23}}}  \\ \\
    \dfrac{\rho_{(12)(23)}}{\sqrt{\tau_{12} \tau_{23}}}      & \quad \dfrac{\rho_{(13)(23)}}{\sqrt{\tau_{13} \tau_{23}}}              & \quad \dfrac{1}{\tau_{23}} \\
  \end{array}
\right).
\end{equation}

This matrix includes two types of hyperparameters: $\tau_{ij}$, the marginal precision of the random effects associated to transition $i \rightarrow j$, and $\rho_{(ij)(i'j')}$, the correlation between the random effects on transitions $i \rightarrow j$ and $i' \rightarrow j'$.

Note that for identifiability reasons we fix $\tau=1$ for the multivariate version of the Leroux model, so that the covariance matrix $\boldsymbol{\Sigma}_{between}$ becomes the one capturing dispersion.

\subsection{Bayesian inference and prior specification}\label{subsec:prior}

A Bayesian approach based on the integrated nested Laplace approximation (INLA) \cite{rue inla}  has been considered to estimate the posterior distribution   of all the quantities of interest of the model. Bayesian inference combines  prior knowledge of all unknown parameters and hyperparameters $\boldsymbol{\theta}$ of the model in probabilistic terms throught the prior distribution with the likelihood function obtained from data $\mathcal D$ by means of the Bayes' theorem to derive the joint posterior distribution $\pi(\boldsymbol{\theta}, \boldsymbol{\psi} \mid \mathcal D)$  of the  parameters and hyperparameters $\boldsymbol{\theta}$  and random effects $\boldsymbol{\psi}$. As models get more complex, it is harder to find an analytic expression for those posterior distributions and computational methods are required to approach them. The most popular procedures are Markov chain Monte Carlo (MCMC) methods \cite{mcmc} which, in most cases, imply large computational times to ensure convergence of the estimations. Alternatively, INLA is a fast and accurate option. It uses  Laplace approximations to obtain the approximated marginal posterior distribution of the parameters, hyperparameters and  latent terms of the sampling model. Survival models, including Cox proportional hazards models, can be adapted and implemented in INLA because they can be expressed in terms of Gaussian Markov random field (GMRF) models \cite{martino}. In particular, competing risks models \cite{niekerk competing}, and illness-death models as an extension of them, can be approached using INLA. It also allows the inclusion of gaussian random effects in the regression term of the Cox proportional hazards model, and thus the proposed spatial illness-death model is naturally approachable with INLA.

As just discussed, we need to complete the Bayesian model with a prior distribution for all parameters and hyperparameters $\boldsymbol{\theta}$ of the sampling model. We have considered a framework of prior independence between the different elements in $\boldsymbol{\theta}$. The shape parameters $\alpha_{ij}$ of the baseline hazard functions in (\ref{eqn:Cox}) were assumed to follow a penalized complexity prior (PC prior) as described in INLA documentation (See \texttt{inla.doc("pc.alphaw")} for a detailed definition). Those PC priors consider an exponential as the base model, i.e. a Weibull model with $\alpha_{ij}=1$, and penalize the departure from this exponential model. The more general Weibull model would be preferred only if there exists enough evidence supporting it \cite{niekerk weibull}. Meanwhile, the scale parameter for the baseline transition intensities, $\lambda_{ij}$, has not a prior by itself but through considering an intercept $\beta_{ij,0}=log(\lambda_{ij})$ which follows, as well as the regression coefficients $\beta_{ij,l}$, a gaussian distribution with mean 0 and precision $0.001$.

The covariance matrix including correlation between transitions, $\boldsymbol{\Sigma}_{between}$, was assumed to follow an inverse Wishart distribution, or equivalently, a Wishart distribution for the precision matrix $\boldsymbol{\Sigma}_{between}^{-1}$. The Wishart distribution is a multivariate generalization of the gamma distribution \cite{gelman bda}. It is specially relevant when modelling correlated normal random effects as it is a conjugated prior distribution  for the precision matrix from multivariate normal distributions, being the most common choice when inferring covariance matrixes. In our case, the prior values  for the Wishart parameters were those provided by default in the INLA specification of the model for correlated random effects, i.e, $\boldsymbol{\Sigma}_{betwen}^{-1} \sim \textrm{Wishart}_3(\nu=7, \boldsymbol{\mathrm{R}}=\boldsymbol{\mathrm{I}})$, where $\nu=7$ are the degrees of freedom. Wishart distribution with the identity as the scale matrix is typically set as a relatively uninformative prior. Note however that several authors have discussed its appropriateness and some alternatives have been proposed. For instance, this prior specification might not be appropriated in the presence of parameters with small variances, resulting in a strongly informative prior distribution \cite{schuurman}. On the other hand, separation-strategies decompose the covariance matrix into variance and correlation components, being it possible to specify separate priors for each component. Correlation matrix obtained after this separation may be assumed to follow an inverse Wishart distribution\cite{omalley}, or a Lewandowski-Kurowicka-Joe \cite{lewandowski}, for instance. Meanwhile, many positive priors can be set for variances such as truncated-normal, half-normal, half-Cauchy or uniform distributions \cite{gelman bda}.

A non-informative uniform prior, $\textrm{U}(0,1)$ was assumed for the mixture parameter $\gamma$ of the  Leroux modelling.   To define the multivariate Leroux model for random effects we used the \texttt{rgeneric} latent effect. Using this mechanism, latent effects can be implemented in INLA via R \cite{gomez-rubio}. Despite they are not specifically applied to survival models, some multivariate versions of random effects have already been defined using this method, such as intrinsic multivariate CAR latent effects, and collected in the \texttt{INLAMSM} package for \texttt{R} \cite{palmi}.

\subsection{Posterior outcomes}\label{subsec:outcomes}

The posterior distribution $\pi(\boldsymbol{\theta}, \boldsymbol{\psi} \mid \mathcal D)$  contains all updated information on the random behaviour of the illness-death population.   Nevertheless, it provides unclear practical  evidence about   the prognostic clinical status for a patient over time. Compound measures such as sojourn times distributions, transition and occupation probabilities and cumulative incidences functions \cite{meira biometrical, touraine} mix the time-evolution information from the illness-death setting as well as the risk-variation among regions. They are specially relevant   in order to gain insight into the clinical setting. From a statistical point of view, posterior inferences of those quantities of interest is straightforward as they are indeed defined as functions of the aforementioned parameters and effects. We introduce some of these  measures of performance and discuss their posterior estimation.

\subsubsection{Sojourn times.}

Sojourn time in state $i$ for an individual living in the site $k$ refers to the time an individual remains in that state without leaving. Possibly, the most interesting sojourn time in illness-death models  corresponds to the initial state. It is defined in terms of the conditional survival function as follows
\begin{align}
S_{1}^{(k)}(t\mid \boldsymbol{\theta}, \boldsymbol{\psi})& =P(T^{(k)}>t \mid \boldsymbol{\theta}, \boldsymbol{\psi}),  \\
& = \mbox{exp}\Big(- \int_{0}^{t}\,(h_{12}^{(k)}(u\mid \boldsymbol{\theta}, \boldsymbol{\psi})+h_{13}^{(k)}(u\mid \boldsymbol{\theta}, \boldsymbol{\psi}))\,\mbox{d}u \Big) \nonumber
\end{align}
\noindent where $T^{(k)}=\mbox{min}\{T_{12}^{(k)}, T_{13}^{(k)}\}$. Because
sojourn time in state $1$ depends on $(\boldsymbol{\theta}, \boldsymbol{\psi})$, the subsequent posterior distribution
$\pi(S_{1}^{(k)}(t\mid \boldsymbol \theta, \boldsymbol \psi)\mid \mathcal D)$, $\forall t$, can be easily approximate from a simulated sample of the posterior  $\pi(\boldsymbol{\theta}, \boldsymbol{\psi} \mid \mathcal D)$.

\subsubsection{Transition and occupation probabilities.}

Transition probabilities depend on the parameters through the subsequent hazard functions according to (\ref{trans_prob2}). Therefore, their posterior distribution $\pi(p_{ij}^{(k)}(s,t\mid \boldsymbol{\theta}, \boldsymbol{\psi} \mid \mathcal D)$ associated with an individual in region $k$ will be also computed from a simulated sample of $\pi(\boldsymbol{\theta}, \boldsymbol{\psi} \mid \mathcal D)$. Occupation probabilities refers to   probabilities associated to the presence of the process in each of the different states at a given time $t$. They can be expressed as transition probabilities $\pi(p_{ij}^{(k)}(0,t\mid \boldsymbol{\theta}, \boldsymbol{\psi} \mid \mathcal D)$, and consequently its posterior distribution could be also approximated  from an approximate sample from $\pi(\boldsymbol{\theta}, \boldsymbol{\psi} \mid \mathcal D)$.

\subsubsection{Cumulative incidence functions.}

Cumulative incidence functions are more frequently used in competing risks environments but  they are also useful for illness-death models, specially when illness is relevant by itself and not only as an intermediate state between the initial state and death. They can be defined equivalently to the competing scenario for survival times $T_{12}^{(k)}$ and $T_{13}^{(k)}$ as follows

\begin{align}
F_{12}^{(k)}(t\mid \boldsymbol{\theta}, \boldsymbol{\psi})= & P(T^{(k)} \leq t, \delta^{(k)}=1  \mid \boldsymbol{\theta}, \boldsymbol{\psi}),\\
F_{13}^{(k)}(t\mid \boldsymbol{\theta}, \boldsymbol{\psi})= & P(T^{(k)} \leq t, \delta^{(k)}=0  \mid \boldsymbol{\theta}, \boldsymbol{\psi}),
\end{align}

\noindent where $\delta^{(k)}$ is the indicator function with value 1 if $T_{12}^{(k)}<T_{13}^{(k)}$ and 0 otherwise. They can be interpreted as the probability at time $t$ of having moved directly from the initial state 1 to state $j$, $j=2,3$, keeping this sense of accumulation as its name suggests. Cumulative incidence regarding the illness state 2 is highly informative because it indicates how many individuals are expected to suffer the illness. It can also be directly compared with the transition probability from state 1 to 2,   which indicates the expected rate of patients who experienced illness and are still alive. Cumulative incidence functions are also expressed in terms of $(\boldsymbol{\theta}, \boldsymbol{\psi})$ as
\begin{equation}
F_{1j}^{(k)}(t\mid \boldsymbol{\theta}, \boldsymbol{\psi})=\int_0^t h_{1j}^{(k)}(s)\exp\Big\{-\int_0^s (h_{12}^{(k)}(u)+h_{13}^{(k)}(u) \textrm{d}u\Big\} \, \textrm{d}s, \,\,j=1,2.
\end{equation}
Consequently, the posterior distribution  of each of  these cumulative incidences,  $\pi(F_{1j}^{(k)}(t\mid \boldsymbol{\theta}, \boldsymbol{\psi})\mid \mathcal D)$, can also be approached by simulated samples of the posterior distribution $\pi(\boldsymbol{\theta}, \boldsymbol{\psi} \mid \mathcal D)$.

\section{A study of recurrent hip fractures in elderly patients}\label{sec:appli}

Clinical settings involving progression of non-terminal diseases, repeated events and populations with a considerable competing risk of death are the main scenarios where multi-state models can be applied. We illustrate here the application of the previous  model  on a study of recurrent hip fracture.

\subsection{The PREV2FO cohort}\label{subsec:prev2fo}

We analyse the PREV2FO cohort, a population-based cohort comprising patients aged 65 years and older discharged after hospitalization for an osteoporotic hip fracture in the Valencia Region (Spain) from January 1, 2008, to December 31, 2015 \cite{llopis}. The Valencia Region is an autonomous community of Spain, with a population of roughly 5 million people (10\% of the Spanish population). The region provides universal healthcare services through the Valencia Health System (VHS) which is an extensive network of public hospitals, primary care centers, and other public resources managed autonomously by the regional government. It is divided in 24 Health Areas, each one corresponding to the administrative area of influence of a public hospital from the VHS.

\begin{figure}[H]
  \centering
  \includegraphics[width=5.5cm]{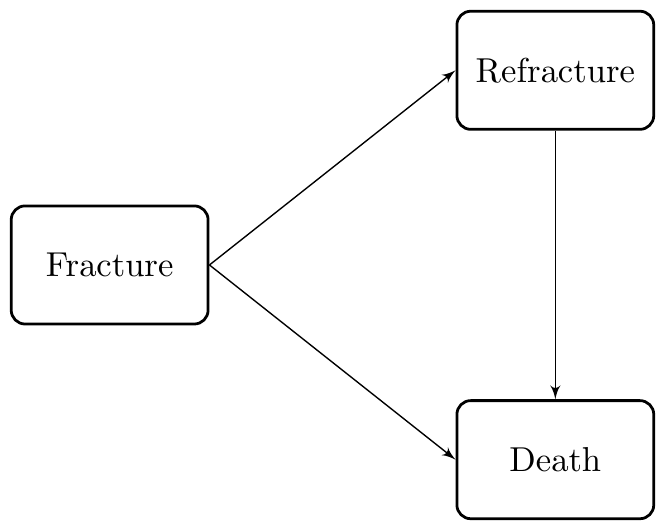}
  \caption{Illness-death model with an initial state of hip fracture, a recurrent hip fracture state and a death state.}\label{diag_hip}
\end{figure}

Patients were followed after the index fracture until death or end of study (December 31, 2016), accounting for recurrent hip fractures during the follow-up period. Figure \ref{diag_hip} shows a diagram  of this process as and illness-death model with an initial state of discharge after a first hip fracture (F), an intermediate state that accounts for discharge after a refracture (R), and   the state of death (D). From a clinical point of view there is a possibility of more than one refracture. We have dismissed this possibility because in our study only a reduced number of the patients suffered from them.

In order to define a basic patient profile we have considered sex,  age at the discharge and the Health Area in which patients were hospitalized as covariates. The study involved $34\hspace*{0.03cm}491$  patients discharged alive after hip fracture, 25$\hspace*{0.03cm}$807 (74.8$\%$)
were women and 8$\hspace*{0.03cm}$684 (25.2$\%$) men. Regarding age, 12.4$\%$ of patients were under
75 years old, 43.6$\%$ between 75 and 85 years old, 40.6$\%$ between 85 and 94 years
old, and 3.4$\%$ were over 95 years old. The mean age at the first  fracture was 83.4
years (IQR: 79.0-88.3). Patients were followed a median time of 5.0 years (IQR:
3.0-7.0 years).

Survival times from state $F$ to $D$, from $F$ to $R$, and from $R$ to $D$ in the Health Area $k$, $k=1, \ldots, 28$, $T^{(k)}_{FD}$, $T^{(k)}_{FR}$ and $T^{(k)}_{RD}$, respectively are modelled by means of a Bayesian spatial Cox proportional hazards model as proposed in previous section.

\subsection{Posterior distribution}\label{subsec:posterior}

We present the  approximated posterior  distribution  sequentially, first the parameters, then the hyperparameters and finally the random effects.   Table \ref{tab_param} summarizes the approximate posterior marginal  distribution of all parameters  of the model. Estimations of the shape parameters of the baseline risk functions, $\alpha_{FD}$, $\alpha_{FR}$, and $\alpha_{RD}$, indicate decreasing hazards over time, specially for the risks of death without and after refracture. $\alpha_{F\!R}$ is closer to 1 which is the threshold which changes the behaviour of the Weibull hazard functions, from increasing to decreasing. Women and men showed no relevant differences in the risk of recurrent hip fracture (E$(\beta_{FR, Woman} \mid \mathcal D)=0.021$), whereas women showed lower mortality risks as compared to men (E$(\beta_{FD, Woman} \mid \mathcal D)<0$, E$(\beta_{RD, Woman} \mid \mathcal D)<0 $). Age was found as a risk factor for refracture and for death without and after refracture.

\begin{table}[h]
\centering
\caption{Summary of the approximate posterior distribution of the parameters from an illness-death model with a multivariate-Leroux model for random effects (I). Transition-related parameters: shape of Weibull distribution and regression coefficients.}
\begin{tabular}{c|c|cccc}
\hline
 Time &\textrm{Parameter}   & Mean & SD & 2.5\% & 97.5\%   \\
  \hline
 From $F$ to $R$ & $\alpha_{F\!R}$  & 0.921 & 0.016 & 0.891 & 0.953    \\
  &$\lambda_{F\!R}$    & 0.028 & 0.005 & 0.018 & 0.040   \\
  & $\beta_{F\!R, \,Woman} $ & 0.021 & 0.050 & -0.076 & 0.119     \\
  & $\beta_{F\!R, \,Age}$  & 0.024 & 0.003 & 0.018 & 0.030    \\
  \hline
  From $F$ to $D$&$\alpha_{F\!D}$  & 0.776 & 0.005 & 0.766 & 0.786      \\
  &$\lambda_{F\!D}$  & 0.335 & 0.054 & 0.238 & 0.460   \\
  &$\beta_{F\!D, \,Woman}$   & -0.510 & 0.017 & -0.543 & -0.477   \\
  &$\beta_{F\!D, \,Age}$   & 0.070 & 0.001 & 0.068 & 0.073 \\
   \hline
From $R$ to $D$& $\alpha_{R\!D}$  & 0.628 & 0.016 & 0.597 & 0.659  \\
&  $\lambda_{R\!D}$  & 0.593 & 0.131 & 0.374 & 0.897 \\
&  $\beta_{R\!D, \,Woman}$  &  -0.634 & 0.065 & -0.761 & -0.505   \\
 & $\beta_{R\!D, \,Age}$  & 0.049 & 0.005 & 0.040 & 0.059 \\
\hline
\end{tabular}
\label{tab_param}
\end{table}

\begin{figure}
  \centering
  \includegraphics[width=9cm]{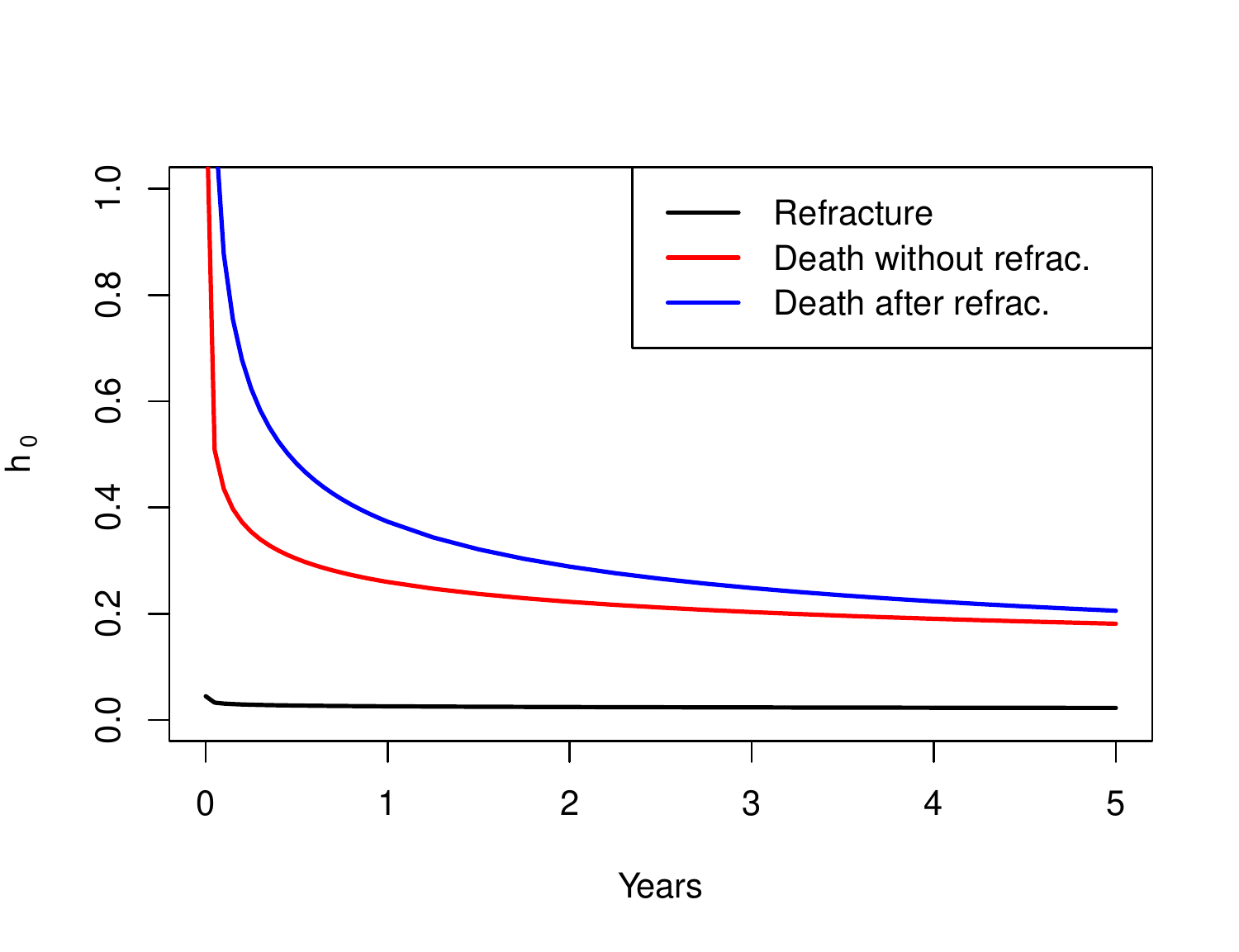}
  \caption{Posterior mean of the baseline hazard function  for each survival transition: from initial hip fracture to recurrent hip fracture, from initial hip fracture to death without refracture and from recurrent hip fracture to death. Horizontal axis indicates years from initial fracture, for transitions from initial fracture to refracture or death without refracture, whilst years from refracture for the transition from refracture to death.}\label{baseline}
\end{figure}

Figure \ref{baseline} shows the posterior expectation of the baseline hazard function associated to each of the three survival times. Note that  risks of death after recurrent hip fracture are higher than those of death without refracture. Transition intensity from fracture to refracture is notably lower than transitions to death. Note that  baseline hazard functions are indeed the hazard functions for the reference values of predictors: average-aged men from a Health Area with random effect equal to 0. Baseline functions suggest higher hazards during the first year, including the hazard of refracture, despite it cannot be appreciated graphically. It results in a sharper increase in the cumulative incidence of those events, as well as greater increases or decreases in the transition probabilities during the initial follow-up.

\begin{figure}
  \centering
  \includegraphics[width=8cm]{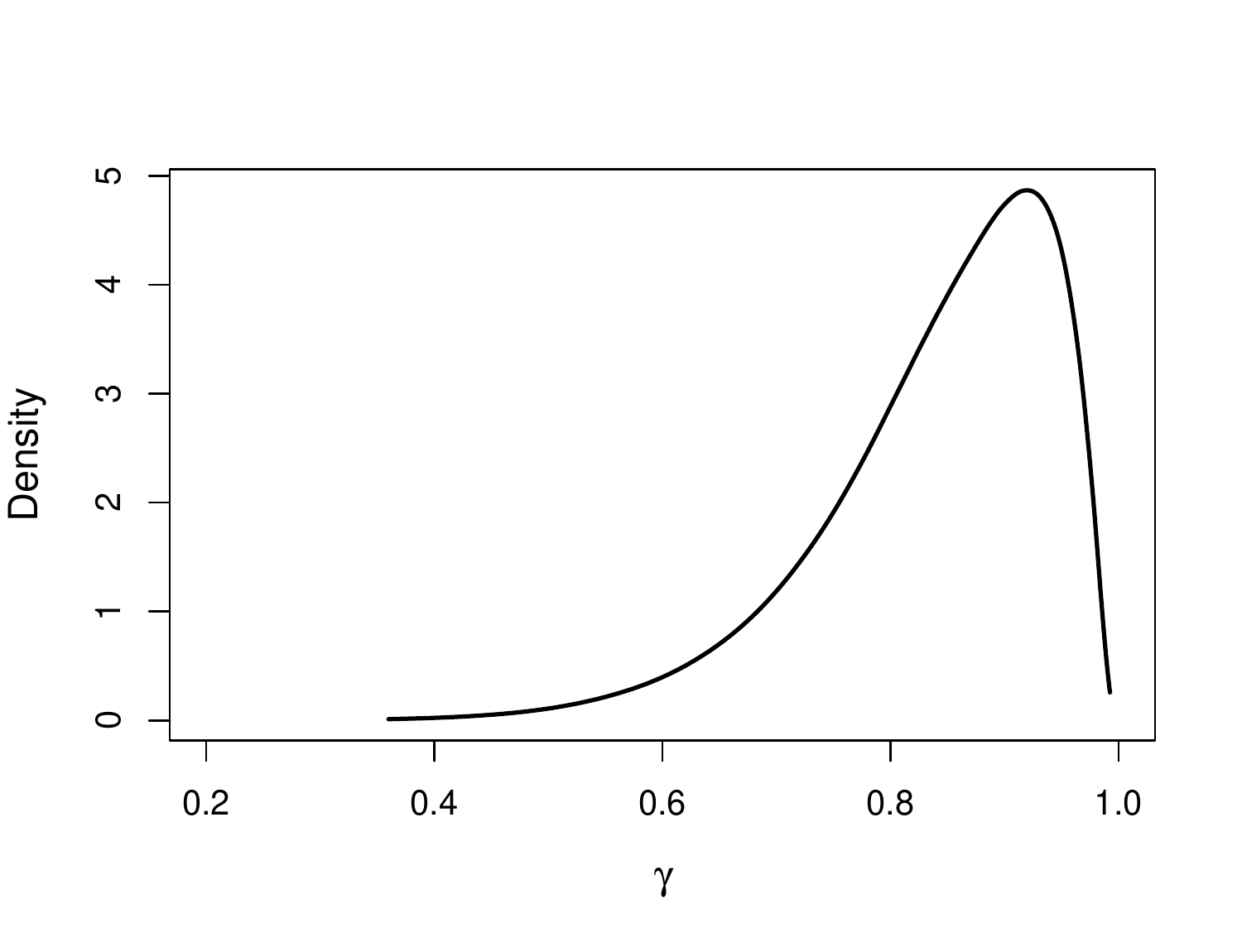}
  \caption{Approximated posterior distribution of the $\gamma$ parameter from an illness-death model with multivariate-Leroux random effects.}\label{gamma}
\end{figure}

Table \ref{tab_param_re} presents a summary of the approximate posterior marginal distribution of the hyperparameters of the spatial illness-death model, all them associated to the  variability   between the  transition survival times and within the different Health Areas in the Valencia Region. The estimation of the parameter $\gamma$ from the Leroux model is  $0.841$ thus indicating that the mixture of an independent scenario and an intrinsic CAR model lends toward the second (Figure \ref{gamma}). A 95\% credible interval excludes lower values suggesting a relevant spatial correlation between areas. Correlation parameters between transitions showed posterior distributions not only including 0 but also zero-centered, which indicates irrelevant correlation parameters thus indicating an uncorrelated scenario. The highest value however was estimated for the correlation between death without refracture and death after refracture, $\rho_{(F\!D)(R\!D)}$, showing a slight correlation between both types of mortality. Uncertainty about random effects is given by precision parameters $\tau$. Higher precision estimations indicate lower variability among random effects. Although the magnitude of the three is very similar, ordered from least to most uncertainty we have random effects on transitions of death without refracture, refracture, and death after refracture.

\begin{table}
\centering
\caption{Summary of the approximate posterior distribution of the hyperparameters from the illness-death model estimated with a multivariate-Leroux model for the spatial random effects. $\gamma$ parameter from the Leroux model, precision of the random effects, and correlation between transitions times.}
\begin{tabular}{rrrrr}
  \hline
 \textrm{Parameter}   & Mean & SD & 2.5\% & 97.5\%  \\
  \hline
  $\gamma$ & 0.841 & 0.101 & 0.591 & 0.973 \\
  $\tau_{F\!R}$ & 14.257 & 4.620 & 7.197 & 25.185 \\
  $\tau_{F\!D}$ & 19.896 & 5.595 & 10.915 & 32.737 \\
  $\tau_{R\!D}$ & 11.743 & 4.181 & 5.386 & 21.625 \\
  $\rho_{(F\!R)(F\!D)}$ & -0.044 & 0.181 & -0.388 & 0.315 \\
  $\rho_{(F\!R)(R\!D)}$ & -0.076 & 0.178 & -0.415 & 0.275 \\
  $\rho_{(F\!D)(R\!D)}$ & 0.109 & 0.164 & -0.217 & 0.423 \\
   \hline
\end{tabular}
\label{tab_param_re}
\end{table}

\begin{figure}
  \centering
  \includegraphics[width=15cm]{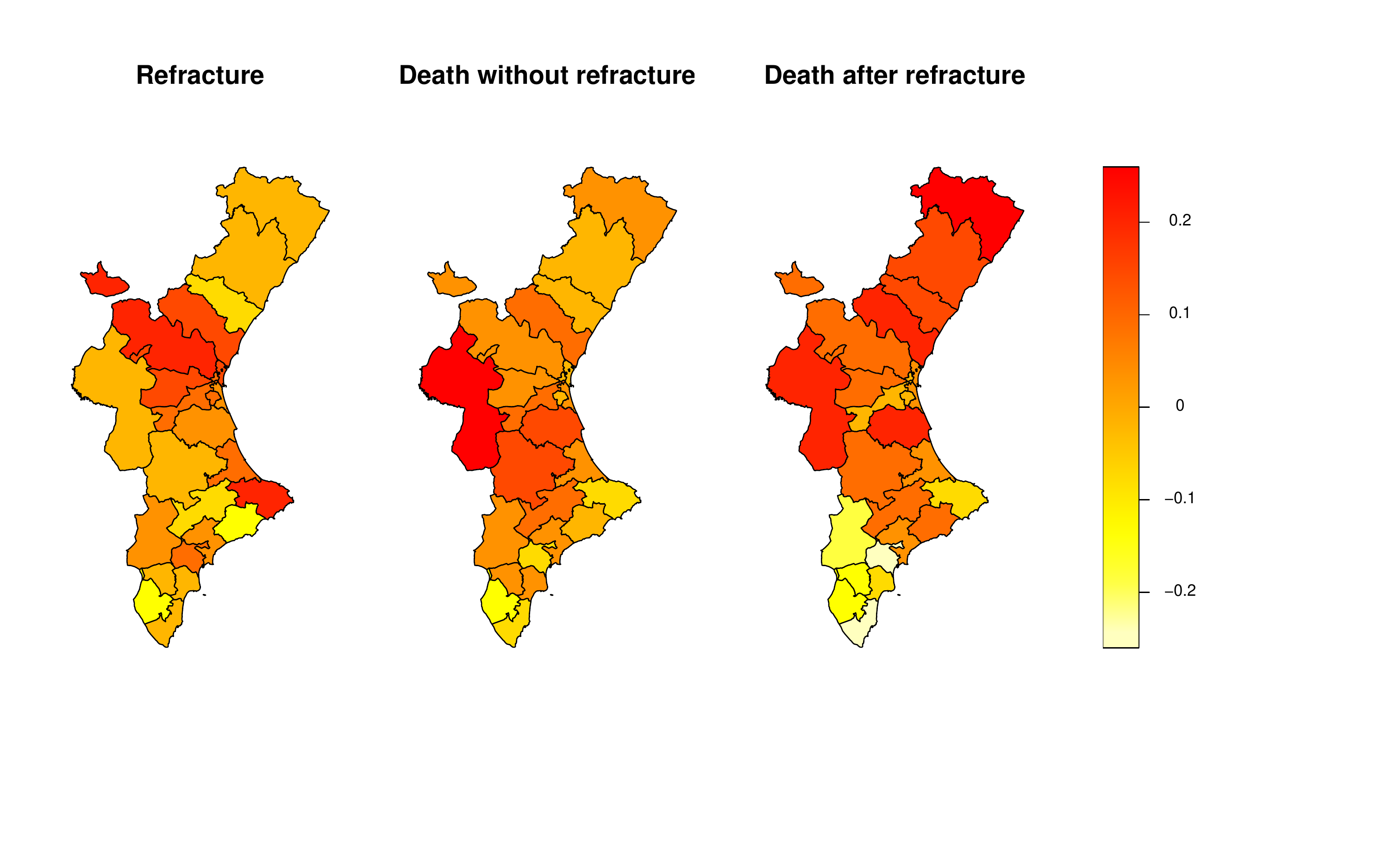}
  \vspace{-2cm}
  \caption{Posterior  mean of the region-specific random effects by Health Area of the Valencia Region, from an illness-death model with multivariate-Leroux random effects.}\label{maps_re}
\end{figure}

Figure \ref{maps_re} displays the posterior mean of the random effects associated to each transition time and Health Area of the Valencia Region. Health Areas coloured red indicate a higher risk of experiencing the event of interest compared to the overall average for all areas. Areas shaded in yellow indicate the opposite. The random effects associated with the three survival times of the illness model from the same Health Area do not always behave the same. We can observe some areas with positive random effects in the three survival times considered, but also some cases where the effects show negative relationships. There are some particular areas with a particular spatial  pattern. This is the case of Requena-Utiel (the most western Health Area) and Denia (located at the cape in the east of the Valencia Region). The first shows lower risk of recurrent hip fracture and higher risks of death without and after refracture. The latter shows the opposite scenario, higher risk of refracture and lower mortality. Both cases illustrate negative association between the risk of refracture and mortality, whilst positive association between both risks of death.

\subsection{Outcome measures of the hip fracture process}\label{subsec:measures}

The examination of the raw estimations provided by the posterior distribution $\pi(\boldsymbol{\theta}, \boldsymbol{\psi} \mid \mathcal D)$ contains full information about the differences in the risk of each outcome. Nevertheless, it provides unclear evidence about which will be the prognostic for a patient with a hip fracture in each particular Health Area or what would be the general evolution of the survival time transitions in the target population. Information regarding time-evolution from the illness-death setting  and the variation in the risk among Areas are indeed combined in posterior distributions for cumulative incidences and for transition probabilities.

\begin{figure}
  \centering
  \includegraphics[width=14cm]{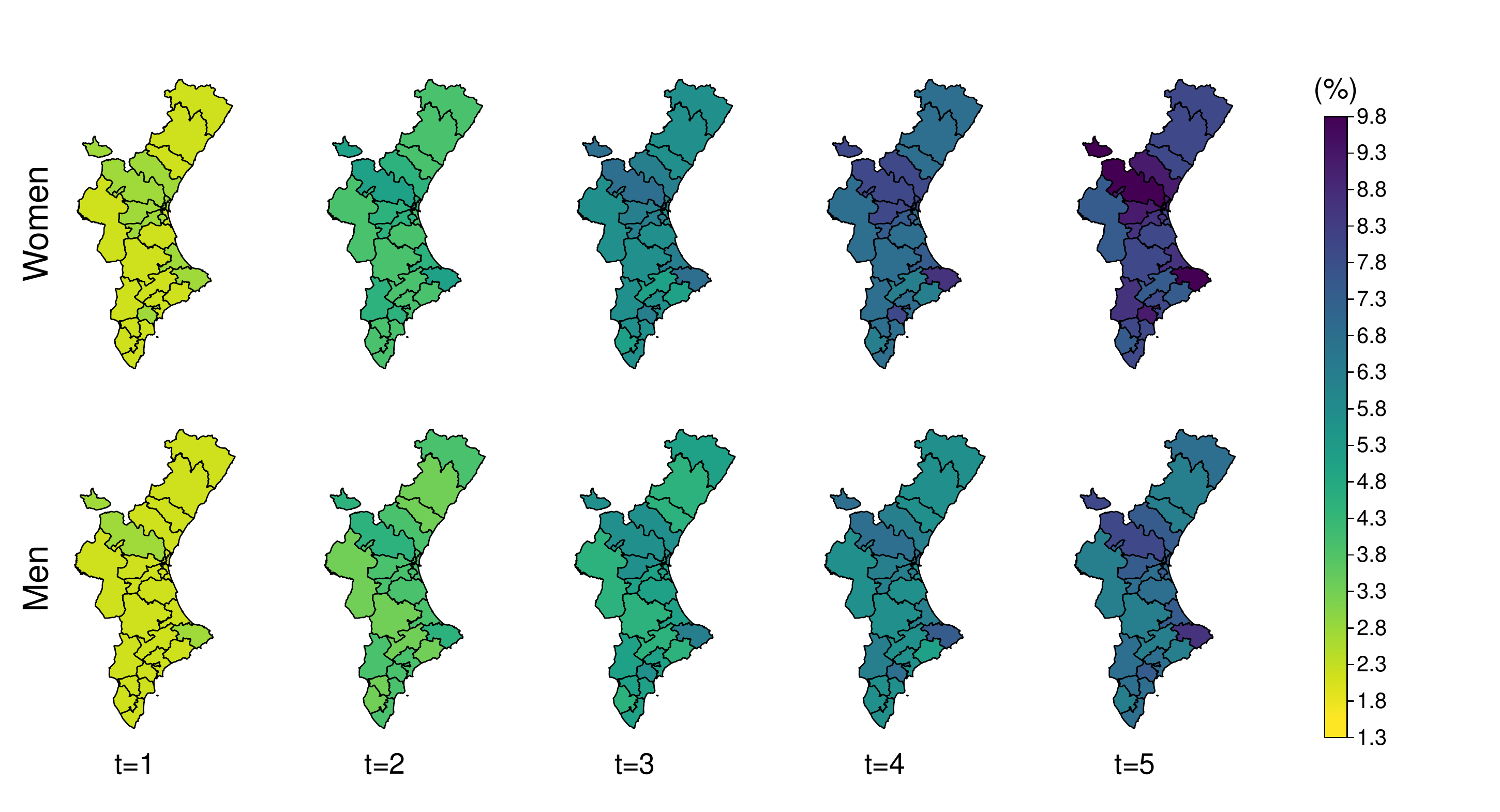}
  \caption{Posterior mean of the cumulative incidence of refracture in 80-year-old women and men at $t=1,2, \ldots,5$ years after a first fracture by Health Area.}\label{maps_CI}
\end{figure}

The cumulative incidence of a hip refracture at time $t$ can be interpreted as the probability of having a hip fracture at a time  before $t$ without having died before that time, as death plays a major role censoring refracture. Note that  higher risk of death leads to the observation of fewer refractures. Therefore, two Areas with the same risk of refracture could show different incidences of refracture depending on the risk of death. Figure \ref{maps_CI} shows the posterior mean of the cumulative incidence of refracture for  80-year-old women and men  in the different Health Areas of the Valencia Region at $t=1,2, \ldots,5$ years after a first hip fracture. In broad terms, higher incidences of recurrent hip fracture are estimated for those regions with higher risk of refracture as it is expected. Those differences become more visible after some years from the initial fracture. The Health Area of Requena-Utiel (the most western region) shows a particular low incidence despite its not so low risk, which can be related to being the region with the highest risk of death without refracture. Men show lower incidences of refracture due to their increased risk of death, as we found no differences in the risk of refracture compared with women. Men reach the same incidence values than women with a delay of 1-2 years approximately.

\begin{figure}
  \centering
  \includegraphics[width=14cm]{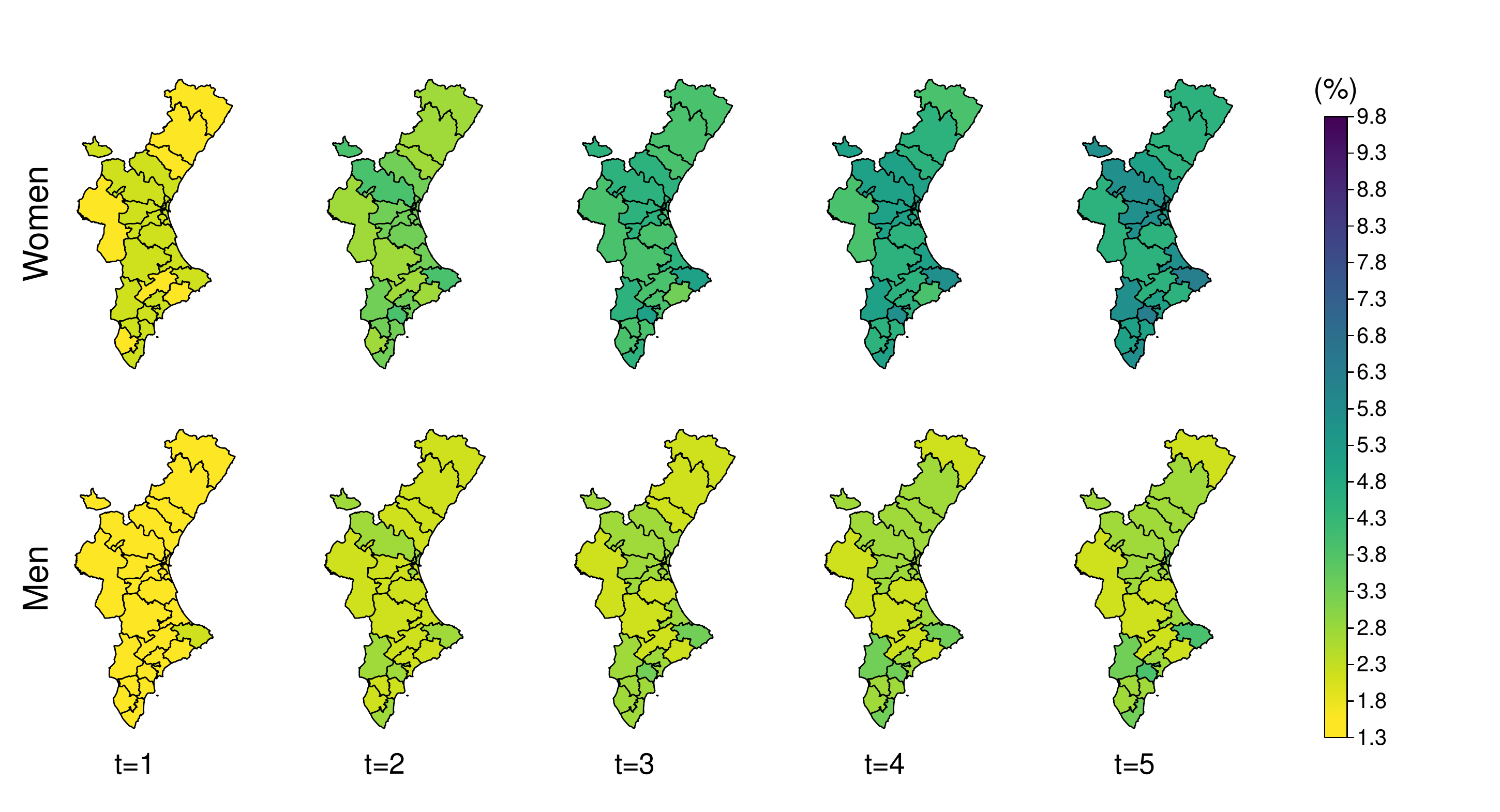}
  \caption{Posterior mean of the transition probability from fracture to refracture ($p_{F\!R}$) in 80-year-old women and men at $t=1,2, \ldots,5$ years after a first fracture by Health Area.}\label{maps_pFR}
\end{figure}

\begin{figure}
  \centering
  \includegraphics[width=14cm]{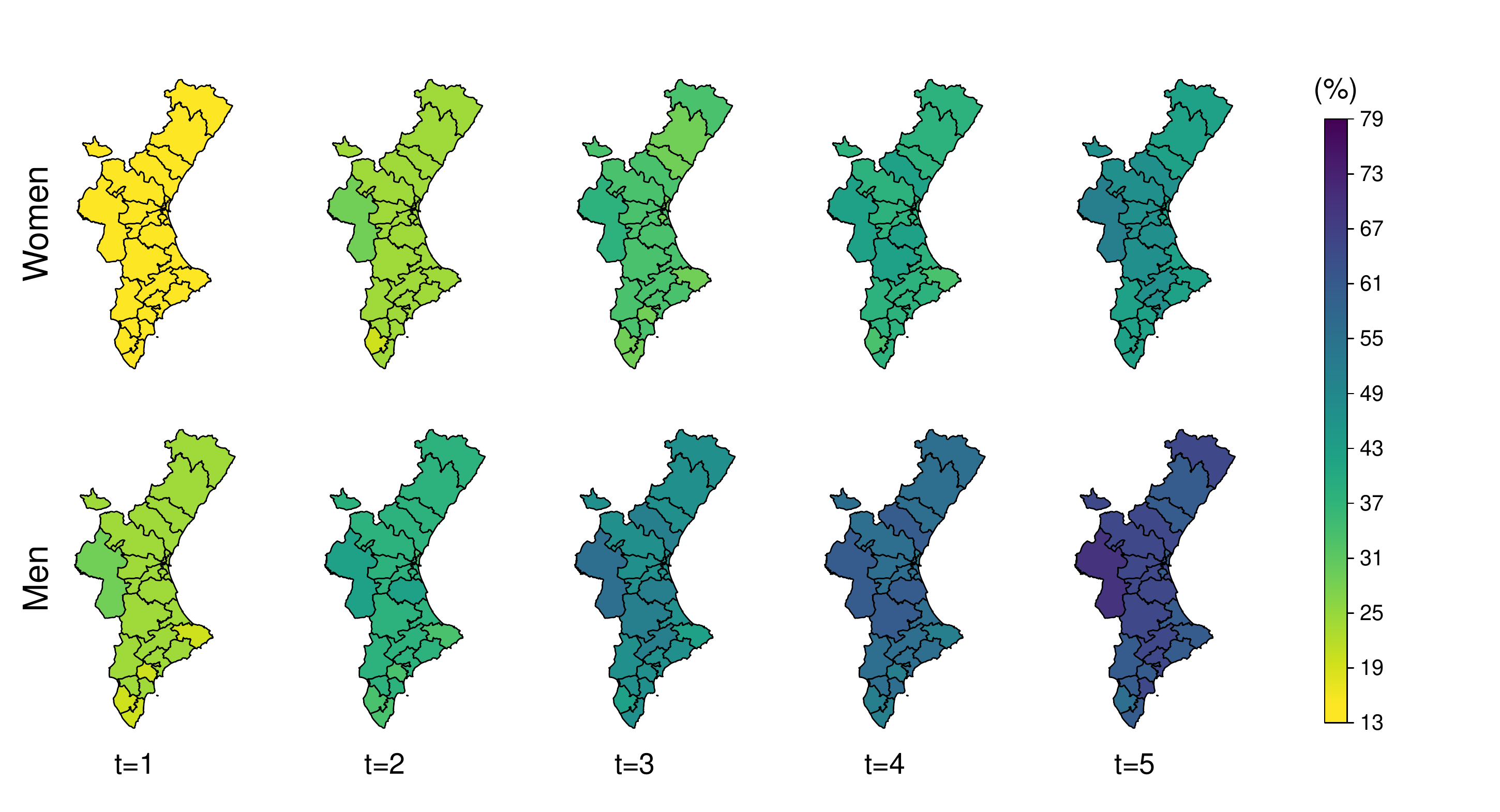}
  \caption{Posterior mean of the total-death probability ($p_{F\!D}$) in 80-year-old women and men at $t=1,2, \ldots,5$ years after a first fracture by Health Area.}\label{maps_pFD}
\end{figure}

\begin{figure}
  \centering
  \includegraphics[width=14cm]{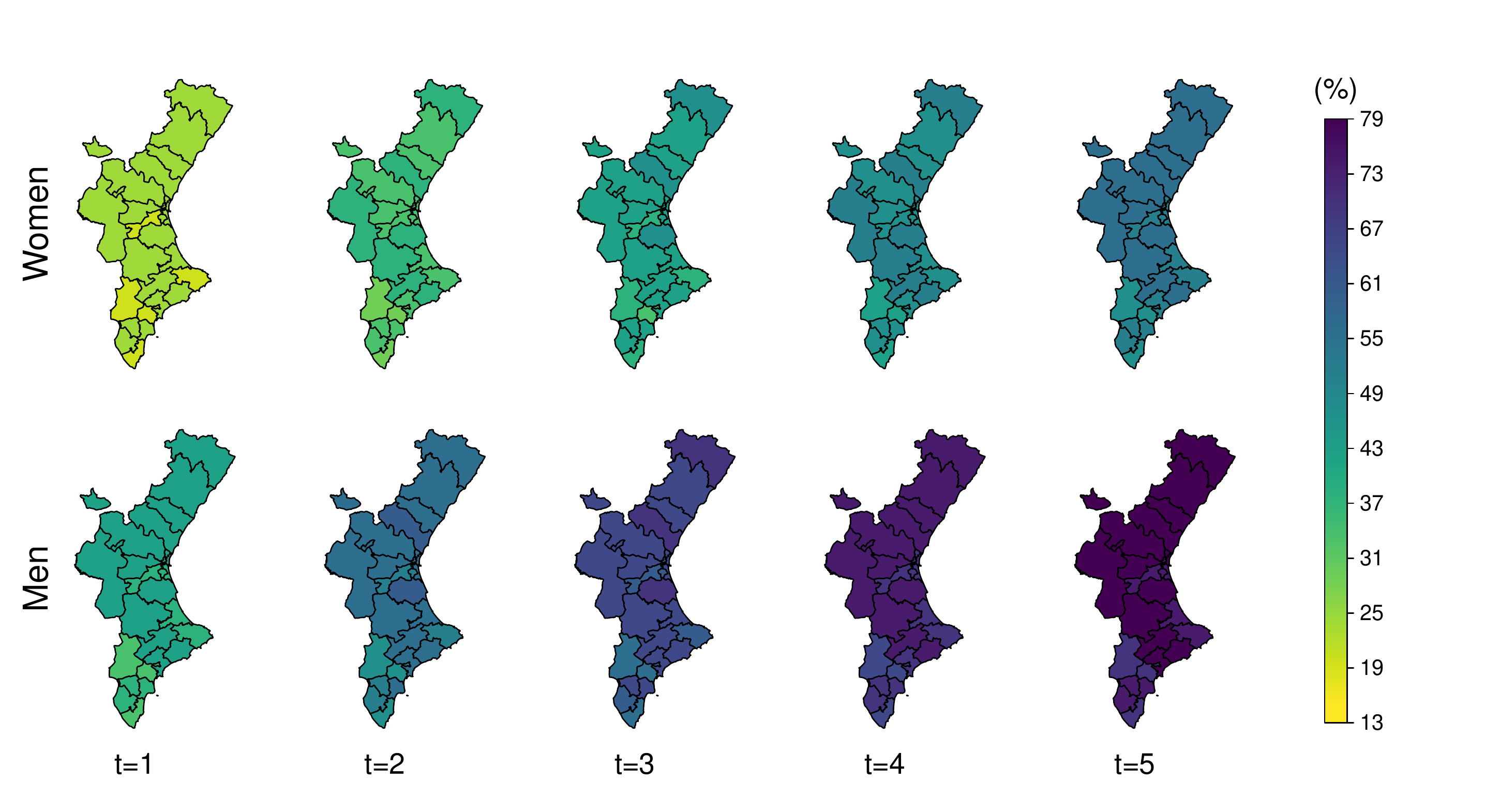}
  \caption{Posterior mean of the probability of death after refracture ($p_{R\!D}$) in 80-year-old women and men at $t=1,2, \ldots,5$ years after a refracture by Health Area.}\label{maps_pRD}
\end{figure}

Regarding  transition probabilities from fracture to refracture (Figure \ref{maps_pFR}), they are  also higher for women, as they are more likely to experience a refracture than men. It shows an increasing trend during 2 and 4 years from the initial fracture for men and women, respectively. After this time, the probability of being refractured and alive remains stable, as the number of patients at risk of refracture decreases and the mortality after refracture offsets the number of new refractures.

Mortality is higher in men for both, total mortality and after recurrent hip fracture only. Women approximately reach at 4 years the same mortality than men at 2 years (Figure \ref{maps_pFD}). This difference is even higher for death after refracture. Women reach at 5 years after refracture the same mortality rates than men only 2 years after refracture (Figure \ref{maps_pRD}).

The number of patients who die after a refracture represent a low fraction of the total mortality. Cumulative incidence of refracture indicates that less than 10\% of women experience a refracture 5 years the initial fracture (even lower in men). Thus, the spatial pattern of the total-death probability (Figure \ref{maps_pFD}) is similar to that showed by the random effects on the risk of death without refracture (Figure \ref{maps_re}).

Probability of death is higher for those patients with a recurrent hip fracture. Mortality one year after refracture is similar to that expected two years after the initial fracture. Its spatial pattern is also different with respect to that shown by total-mortality, and is identical to that shown by the respective random effects on transition from refracture to death in Figure \ref{maps_re}. This is due to the fact that the probability of death after refracture is the only one which depends exclusively on one transition intensity, in particular, the transition intensity from refracture to death.

\section{Discussion}\label{sec:discu}

The potential and usefulness of illness-death models are highly increased after combining it with spatial information modelled by multivariate random effects associated to a set of spatial units. Differences among regions can be studied under this joint framework, in addition to progression of individuals over time. Despite correlation between transitions was not observed to be relevant in our real-world study, the possibility of modelling it jointly with the spatial correlation from a model based on a neighbourhood structure, such as the Leroux model, defines an extensive number of options depending on the needs of the clinical framework. Moreover, one might be interested in using more complex multi-state models with additional states and transitions. It would be analogous as the concept throughout our work regarding how random effects are included in transition intensities is quite general and thus, it is not restricted to illness-death models.

The usage of the integrated nested Laplace approximation (INLA) is another strong point of our work. Computational time is highly reduced compared to MCMC methods, and the introduction of gaussian random effects in the regression terms is something natural in INLA. However, it is not a popular choice when assessing multi-state models yet, although this in the near future might change given its benefits. Multi-state modelling in INLA remains unexplored and further work regarding this issue would be of high interest.

Finally, assessing transition probabilities and cumulative incidences instead of only analysing the estimations of random effects on transition intensities provides a deeper understanding of the clinical or epidemiological problem. In fact, risk assessment alone seldom provides predictive information regarding the prognostic of a patient given some specific characteristics. Meanwhile, trajectories expressed in terms of probabilities are dynamic and interpretable outcomes which are essential to reach an individualized care and that can be determinant to clinicians and policy makers.

\subsection*{Declaration of competing interests}
The author(s) declared no potential conflicts of interest with respect to the research, authorship, and/or publication of this article.

\subsection*{Funding}
The author(s) disclosed receipt of the following financial support for the research, authorship, and/or publication of this article: This work has been funded by Instituto de Salud Carlos III (ISCIII) through the projects [PI14/00993, PI18/01675], ``RD16/0001/0011 - Red Temática de Servicios de Salud Orientados a Enfermedades Crónicas (REDISSEC)", and ``RD21/0016/0006 - Red de Investigación en Cronicidad, Atención Primaria y Promoción de la Salud (RICAPPS)", and co-funded by the European Union. FLC was funded by Instituto de Salud Carlos III (ISCIII) [grant number FI19/00190], and co-funded by the European Union. CA was partially funded by Ministerio de Ciencia e Innovaci\'on (MCI, Spain) [grant number PID2019-106341GB-I00].

\end{document}